# A Re-examination of the Census Bureau Reconstruction and Reidentification Attack

Krishnamurty Muralidhar, Price College of Business, University of Oklahoma, Norman, OK
krishm@ou.edu

**Abstract**: Recent analysis by researchers at the U.S. Census Bureau claims that by reconstructing the tabular data released from the 2010 Census, it is possible to reconstruct the original data and, using an accurate external data file with identity, reidentify 179 million respondents (approximately 58% of the population). This study shows that there are a practically infinite number of possible reconstructions, and each reconstruction leads to assigning a different identity to the respondents in the reconstructed data. The results reported by the Census Bureau researchers are based on just one of these infinite possible reconstructions and is easily refuted by an alternate reconstruction. Without definitive proof that the reconstruction is unique, or at the very least, that most reconstructions lead to the assignment of the same identity to the same respondent, claims of confirmed reidentification are highly suspect and easily refuted.

## Introduction

According to the declaration by Dr. John Abowd, Chief Scientist and Associate Director for Research and Methodology at the United States Census Bureau, the disclosure prevention procedures used in the 2010 Census did not prevent the ability of an adversary to identify the Census respondents and results in the confirmed reidentification of as many as 179 million respondents to the 2010 Census. (Abowd, 2021a, p. 12). The complete procedure used by the Census Bureau (hereafter, REID) consisted of three steps:

(1) Reconstruct microdata (individual level data) for all respondents in the US by using publicly available Census data.
(2) Link the reconstructed microdata to a commercial database using (Age, Sex) and assign a name and address to the reconstructed microdata records.
(3) Compare the enhanced microdata with the original Census data to confirm the identity of the respondents.

The Census collects both individual and household level data. The original data that is gathered is edited for errors and any other issues and the Census Edited File (CEF) is created. All personally identifiable information is removed and replaced with a unique identifier, the Protected Identification Key. Statistical disclosure limitation procedures (privacy protection measures) are applied to CEF which results in the creation of the Hundred Percent Detail File (HDF). All Census publications are produced from HDF.

The Census data released to the public are released in two categories, personal and household. No linkage between the two categories is provided. The personal level data that is released to the public consists of (Age, Sex, Race, and Ethnicity). Household level data is similar but has additional information (number of individuals in household, relationship to the householder) and the data release also provides additional information on the Age variable (average and median).

The Census releases data at different geographic levels: nation, state, county, tract, block group, and block. The final three are census-defined constructs and do not necessarily correspond to traditional geographic classification. For personal level data, the data at the smaller geographic level is aggregated to the next higher level, that is, the results at the block level are aggregated to block groups, block groups are aggregated to tracts, etc. The multiple tables that are released (Total Population, Sex by Age, Total Races, and others) are all aggregations of the most detailed data release (Age by Sex, by Race, by Ethnicity). The different tables released form the basis of the reconstruction of the respondent microdata.

Every respondent record consists of both personal information and information about how the respondent is related to the householder (the primary individual in the household in whose name the housing unit is owned or rented). Information regarding the relationship variable is only released in the household tables and not as a part of the individual level data. During the reconstruction, the REID team did not recreate the entire record for the respondent, but only the variables released in individual level tables, namely, Age, Sex, Race, Ethnicity (Abowd, 2021a). REID procedure is implemented using two external data files: commercial data sources and CEF as the external data file.

Unfortunately, it is impossible for anyone outside the Census to have access to either of these two files. As far as the commercial data, while the sources of the data have been identified, the accuracy of these specific data files were never identified (Rastogi & O'Hara, 2012). Since the reidentification claims are directly affected by the accuracy of the external data, the accuracy of the external data source must be verified. However, it is very difficult (if not impossible) to recreate this external data file to serve as a comparison. According to Abowd (2021a), using CEF as the external data file is a worst-case scenario since it is the most accurate data that an adversary can have. As a result, the choice of CEF as the external data favors the results and claims of REID. Again, unfortunately, the Census Bureau does not provide access to CEF without special authorization. Thus, we have a situation where it is practically impossible to gain access to the data to verify the results of REID. To overcome this impossible situation, I have chosen to generate a hypothetical CEF based on the characteristics of the available data and as the "external data" file. Hence, as with the use of the true CEF, it is assumed that there is no inaccuracy between the true and external data.

The analysis in this paper is based on data from Tract 5.01, Laramie County, Wyoming (https://data.census.gov/cedsci/advanced). It should be noted that the choice of this tract was simply a matter of convenience. Similar data are available in practically every county in every state in the nation. The tract consists of a total of 148 blocks, 127 occupied blocks, and a total population of 8164. The largest block in the tract has a population of 450 and the smallest occupied block has a population of 1. Table 1 shows the race and ethnicity breakdown for this tract.

|  | White | Black | American Indian or Alaskan Native (AIAN) | Asian | Native Hawaiian or Pacific Islander (NHPI) | Other | Two or More Races (Multiple) |
|---|---|---|---|---|---|---|---|
| Not Hispanic | 6420 | 210 | 56 | 115 | 26 | 14 | 155 |
| Hispanic | 688 | 32 | 20 | 4 | 3 | 304 | 117 |

Table 1. Race breakdown for Tract 5.01, Laramie County, Wyoming

## Reconstruction of Respondent-level Data

The basic premise underlying the entire REID experiment can be summarized by the following statement (Abowd, 2021a, p. 14):

> While the statistical and computer science communities have been aware of this vulnerability since 2003, only over the last few years have computing power and the sophisticated numerical optimization software necessary to perform these types of reconstructions advanced enough to permit reconstruction attacks at any significant scale.

This is incorrect. The Census data files from 2010 (and even 2000) could be used reconstruct the microdata for every respondent in the nation very easily. At the tract level, Census releases tables of count by individual year of age, sex, race, and ethnicity (PCT12A-O). To reconstruct the data at the tract level is just a matter of creating a list of individuals based on the counts provided in the tables. The difference between the description above and the reconstruction in the REID experiment is the level of geography. The reconstruction above is at the tract level while the REID reconstruction is at the block level. At the block level, the Age variable is grouped (except for ages 20 and 21). In addition, other than White respondents, Age by Sex is only provided for the Race category as a whole and breakdown by Ethnicity is not provided. The reconstruction procedure at the block level must be applied twice (first for Hispanic respondents followed by non-Hispanic respondents) for all respondents who are not White.

Within each tract, the reconstruction can be performed independently for each Sex and Age Group (23 in total). My analysis in this paper focuses on Males in the Age Group (25 − 29) in Tract 5.01 in Laramie County, Wyoming. At the tract level, there are a total of 338 respondents in this Age Group in 87 different blocks. From Tables P8 and P9, the tract level race breakdown for this Age Group is shown in Table 2. From Tables P8 and P9 at the block level, the adversary can also create a similar table for each of the 87 blocks (with the exception of Ethnicity for non-White individuals as noted earlier). Table 2 provides information for one such block (Block 4000) in Tract 5.01.

Given the information for Block 4000 in Table 2, reconstructing the individuals in Block 4000 is simply a matter of creating a list as shown in Table 3. The individuals in all other blocks can be similarly reconstructed with the exception that this reconstruction presents age only as a group (25 – 29), rather than individual year of age.

| | | White | Black | AIAN | Asian | NHPI | Other | Multiple |
|---|---|---|---|---|---|---|---|---|
| Tract 5.01 | Not Hispanic | 263 | 13 | 3 | 6 | 1 | 1 | 3 |
| | Hispanic | 28 | 0 | 2 | 0 | 1 | 12 | 5 |
| Block 4000 | Not Hispanic | 9 | 0 | 0 | 1 | 0 | 0 | 0 |
| | Hispanic | 3 | 0 | 0 | 0 | 0 | 0 | 0 |

Table 2. Race breakdown for ages (25 – 29), Tract 5.01, Laramie County, Wyoming

| Tract | Block | Sex | Race | Ethnicity | Age |
|---|---|---|---|---|---|
| 5.01 | 4000 | Male | White | Not Hispanic | (25 – 29) |
| 5.01 | 4000 | Male | White | Not Hispanic | (25 – 29) |
| 5.01 | 4000 | Male | White | Not Hispanic | (25 – 29) |
| 5.01 | 4000 | Male | White | Not Hispanic | (25 – 29) |
| 5.01 | 4000 | Male | White | Not Hispanic | (25 – 29) |
| 5.01 | 4000 | Male | White | Not Hispanic | (25 – 29) |
| 5.01 | 4000 | Male | White | Not Hispanic | (25 – 29) |
| 5.01 | 4000 | Male | White | Not Hispanic | (25 – 29) |
| 5.01 | 4000 | Male | White | Not Hispanic | (25 – 29) |
| 5.01 | 4000 | Male | White | Not Hispanic | (25 – 29) |
| 5.01 | 4000 | Male | White | Hispanic | (25 – 29) |
| 5.01 | 4000 | Male | White | Hispanic | (25 – 29) |
| 5.01 | 4000 | Male | White | Hispanic | (25 – 29) |
| 5.01 | 4000 | Male | Asian | Not Hispanic | (25 – 29) |

Table 3. Reconstructed records for Block 4000 (with Age Groups only)

For all Males in Age Group (25 – 29) in Tract 5.01, individual year of age breakdown by (Race and Ethnicity) can be obtained from PCT12A-O and is reconstructed below in Table 4:

| Ethnicity | Age | White | Black | AIAN | Asian | NHPI | Other | Multiple |
|---|---|---|---|---|---|---|---|---|
| Not Hispanic | 25 | 39 | 1 | 0 | 0 | 0 | 0 | 1 |
| | 26 | 53 | 7 | 1 | 4 | 0 | 1 | 1 |
| | 27 | 56 | 2 | 0 | 1 | 0 | 0 | 0 |
| | 28 | 57 | 2 | 1 | 1 | 1 | 0 | 1 |
| | 29 | 58 | 1 | 1 | 0 | 0 | 0 | 0 |
| Hispanic | 25 | 3 | 0 | 1 | 0 | 0 | 5 | 1 |
| | 26 | 6 | 0 | 0 | 0 | 0 | 1 | 0 |
| | 27 | 6 | 0 | 0 | 0 | 0 | 1 | 0 |
| | 28 | 9 | 0 | 0 | 0 | 1 | 1 | 2 |
| | 29 | 4 | 0 | 1 | 0 | 0 | 4 | 2 |

Table 4. Individual year of age by (Race and Ethnicity) for Tract 5.01

This reconstruction can be performed independently for each Sex and Age Group in the tract. This is precisely why the reconstruction problem "is massively parallel in tracts" (Abowd 2018, p. 16). Note that the values for (Sex, Race, Ethnicity) reconstructed for every individual in every block in Tract 5.01 (like Block 4000 in Table 3) will *always* satisfy all the additivity constraints for these three variables at the tract level. The only missing variable in the reconstructed data is the individual year of age. *Hence, the entire reconstruction problem reduces to one of assigning individual year of age values at the tract level for a given (Sex, Race, Ethnicity) to the respondents in the blocks with the same (Sex, Race, Ethnicity).*

For the purposes of illustration, consider the reconstruction of (Male, Black, non-Hispanic) respondents in the Age Group (25 – 29) in Tract 5.01. Table 5 shows the distribution of the 13 (Male, Black, non-Hispanic) respondents in the (25 – 29) Age group in the blocks in Tract 5.01.

| Block | 3014 | 3017 | 3019 | 3021 | 4002 | 4003 | 4012 | 4021 | 4026 |
|---|---|---|---|---|---|---|---|---|---|
| Respondents in (25 - 29) Age Group | 1 | 1 | 1 | 1 | 1 | 1 | 4 | 1 | 2 |

Table 5. Number of (Male, Black, non-Hispanic) respondents in different blocks in Tract 5.01

The REID approach to the reconstruction of individual year of age for these 13 respondents is to express the problem as a system of linear equations to find individual year of age assignment at the block level which satisfies individual year of age frequencies at the tract level. This system of linear equations is solved using Gurobi optimization software. The purpose of optimization software is to find the best possible solution (optimal) from among many solutions that satisfy the mathematical equations (feasible), where the best possible is evaluated based on the objective function. In some cases, there are multiple optimal solutions, but usually only a few. The reconstruction problem does not have an objective function (there is no reason to treat one reconstruction as being superior to any other), and every feasible solution is an acceptable solution. Hence, the number of potential solutions remains very large.

Even for this small group of 13 individuals, the number of possible solutions runs in the hundreds of thousands. The single age 25 value can be assigned to any one of the 13 respondents in the block in 13 different ways. The seven age 26 values can be assigned to the remaining 12 respondents in 792 different ways. The two age 27 values can be assigned to the remaining five respondents in 10 different ways. The two age 28 values can be assigned to the remaining three respondents in three different ways. Finally, there is only way to assign the single age 29 value to the single remaining respondent. In total, there are $(13 \times 792 \times 10 \times 3 \times 1 =)$ 308,880 different assignments of age values across this small group of individuals. Every one of these assignments is a feasible solution to the system of linear equations representing (Male, Black, non-Hispanic) respondents in Age Group (25 – 29) in Tract 5.01. This process can then be repeated for each Age Group in each (Race, Ethnicity) combination for each Sex in each Tract.

The reconstruction reduces to the problem of assigning individual year of age at the block level while preserving the frequency of the respective age at the tract level. A simpler way to achieve this is to create a *vector* of individual years of age with the same frequency as at the tract level, *randomly* sort this vector, and *assign* them to individuals in each block. Every random sort of the age vector satisfies the age frequencies (1, 7, 2, 2, 1 for ages 25, 26, 27, 28, 29, respectively) at the tract level and represents a feasible solution to the system of linear equations. Table 5 shows five different individual year of age assignments for the (Male, Black, non-Hispanic) individuals in Tract 5.01.

| Block | Respondent | I | II | III | IV | V |
|-------|-----------|-----|-----|-----|-----|-----|
| 3014 | 1 | 25 | 27 | 26 | 28 | 29 |
| 3017 | 1 | 26 | 26 | 27 | 26 | 25 |
| 3019 | 1 | 27 | 27 | 28 | 26 | 26 |
| 3021 | 1 | 26 | 26 | 26 | 26 | 26 |
| 4002 | 1 | 26 | 26 | 26 | 26 | 26 |
| 4003 | 1 | 26 | 28 | 29 | 26 | 28 |
| 4012 | 1 | 29 | 26 | 26 | 29 | 26 |
| 4012 | 2 | 27 | 28 | 26 | 27 | 26 |
| 4012 | 3 | 28 | 29 | 26 | 25 | 26 |
| 4012 | 4 | 28 | 26 | 27 | 26 | 26 |
| 4021 | 1 | 26 | 25 | 26 | 27 | 27 |
| 4026 | 1 | 26 | 26 | 25 | 26 | 28 |
| 4026 | 2 | 26 | 26 | 28 | 28 | 27 |

Table 5. Five different reconstructions of age
(Male, Black non-Hispanic, Age group 20 – 25, Tract 5.01, Laramie County, Wyoming)

White, non-Hispanic respondents constitute 263 of the 338 respondents in Tract 5.01, of whom 39 respondents are of age 25. The possible assignments of just the 39 values of age 25 across the 263 respondents is greater than $10^{46}$. The reconstruction of each race and ethnicity combination is performed independently of all others. As a result, the number of feasible assignments is the product of the possible solutions for each race and ethnicity combination. Suffice it to say that the number of alternative reconstructions is practically infinite. *Every one of these infinite reconstructions represents a feasible solution to the system of linear equations representing the data and consistent with all the tables used by the REID team both at the block and tract level.*

**Putative Reidentification**

The Census reidentification attack is explained in detail in Abowd (2021a). The process is as follows:

> Identify the corresponding census block for every address in the source file. Then, looping through all the records in the reconstructed microdata file produced from the reconstruction, find the first record in the source file that matches exactly on block, sex, and age. Once this step is completed, run through the remaining unmatched records from the reconstructed microdata and find the first unmatched record from the source file that matches exactly on block and sex, and matches on age plus or minus 1 year (Abowd, 2021a, Appendix B, p. 7).

When a match is found, it represents a putative identification, and the identification information is harvested from the external data file and appended to the reconstructed data. Abowd (2021b, Table 2) reports a putative identification of 77% (238,175,305 out of 308,745,538).

This procedure was applied to the 10 different reconstructions for Tract 5.01 and the putative identification results are provided in Table 6. These results are consistent with the national level putative identification rate of 77% observed by Abowd (2021b). When these results are viewed independently, these results seem to provide strong support for the REID reidentification results.

| Reconstruction | 1 | 2 | 3 | 4 | 5 | 6 | 7 | 8 | 9 | 10 |
|---|---|---|---|---|---|---|---|---|---|---|
| Putative Identification | 80% | 77% | 81% | 76% | 76% | 78% | 80% | 79% | 79% | 80% |

Table 6. Aggregate putative identification rates for Tract 5.01

These results, however, cannot be viewed independently. Each row of the 338 reconstructed records in this data set *represents a unique individual (as does the reconstructed record of each of the 308,745,538 respondents across the nation)*. The objective of the REID reconstruction and reidentification experiment is to *assign identity to each of these unique individuals*. The ten different reconstructions presented above for the 338 individuals in Tract 5.01 are the result of applying different values of age (from 25 to 29) to these unique individuals. To conclude that matching with the external data results in putative reidentification, *it is necessary that the same record in the reconstructed data is assigned the same identity on every reconstruction*. Assigning different identities to the same reconstructed record on different reconstructions implies uncertainty in the validity of the reidentification. *The fallacy in the REID approach is to treat a single reconstruction as definitive proof of reidentification*.

It is important to note that this analysis can be performed by the adversary with knowledge of the external data (sans race and ethnicity) and the tabular data released by the Census. *Any intelligent adversary would realize that making claims of reidentification based on a single reconstruction can be immediately refuted by the Census Bureau by presenting one of the infinite alternate reconstructions*. Yet valid reidentification based on a single reconstruction is precisely the claim that the REID team is making.

Generally, the extent of agreement between two alternate reconstructions represents a simple measure of the similarity between two reconstructions. Agreement between two reconstructions was computed as the number of respondents for whom the same identity was assigned in both reconstructions. If most of the reconstructions show strong agreement, that would be evidence that conclusions based on multiple reconstructions will not be very different from one another. Analyzing the 10 reconstructions for Tract 5.01 indicates that this is not the case. The average agreement between any two reconstructions is 16% with a minimum of 10% and a maximum of 21%. There is little confidence that any two reconstructions will lead to the same conclusion.

Table 7 shows the number of times the same identity from the external data file was assigned to the same record in the reconstructed data over all 10 reconstructions for all 338 records. If the same identity is assigned to the same reconstructed record every time, it supports the conclusion that reidentification has occurred, which is not the case. *Not a single record was assigned the same identity across all reconstructions (or even nine out of 10 reconstructions)*. Furthermore, only three reconstructed records were assigned the same identity in eight of the 10 reconstructions. All three records are from blocks with a single individual with ages (28, 28, 26). The probability of a random match from assignment of age for these three individuals is 60% and the probability of observing eight matches from 10 reconstructions is close to 12%. Considering that there are 26 blocks with a single individual, the observed value (3) is almost exactly the expected value (3.14). No blocks with more than five individuals ever had the same identity assigned to a reconstructed record more than five times. For the adversary to be confident in the putative identifications, it is necessary that the same identity is assigned to the same individual on all (or at least most reconstructions). These results provide little or no confidence in the putative identification.

| Number of reconstructions where the same identity was assigned | 1 | 2 | 3 | 4 | 5 | 6 | 7 | 8 | 9 | 10 |
|---|---|---|---|---|---|---|---|---|---|---|
| Number of records | 6 | 96 | 96 | 66 | 41 | 14 | 16 | 3 | 0 | 0 |

Table 7. Number of records for which the same identity was assigned

Table 8, which shows the number of different identities assigned to the same record, provides a different perspective but leads to the same conclusion. The most important observation from this table is that, *not one of the reconstructed records had the same identity assigned to it in all 10 reconstructions*. Almost 75% of the reconstructed records were assigned at least four different identities. From the adversary's perspective, the inability to consistently assign the same identity to the same individual across all 10 reconstructions implies that assigning identity based on the reconstructed data is unreliable.

| Number of different identities assigned over 10 reconstructions | 1 | 2 | 3 | 4 | 5 | 6 | 7 | 8 | 9 | 10 |
|---|---|---|---|---|---|---|---|---|---|---|
| Number of records | 0 | 35 | 52 | 90 | 45 | 48 | 28 | 27 | 12 | 1 |

Table 8. Number of identities assigned to the reconstructed records in Tract 5.01

The number of identities assigned is dictated by the size of the block. When the block size is $k$, there can be no more than $(k + 1)$ identities (number of individuals in the block + no match) that can be assigned. This implies that of the 35 records for which only two identities were assigned, 26 of them belonged to a block with a single individual. For *every* block with 10 or more individuals, at least four identities were assigned to each record. In the largest block, *27 of the 29 records were assigned at least six different identities*.

In addition to the aggregate results in Tables 7 and 8, it is also illustrative to analyze results pertaining to a single block. Consider Block 4012 with a total of 10 individuals: four (White, non-Hispanic), four (Black, non-Hispanic), one (AIAN, Hispanic), and one (White, Hispanic) with unique identifiers (RR226 to RR236 for the reconstructed records and 226 to 336 for individuals in the external data). Table 9 shows the assignment of identity for these individuals on the 10 different reconstructions. Table 9 *shows that no reconstructed record is assigned the same identity more than five times. Every reconstructed record is also assigned between four and eight different identities*.

Table 10 shows the race and ethnicity assigned to every individual (identified by the ID) in the external data file from the 10 reconstructions. There is no individual from the external data who is assigned the same race and ethnicity across all reconstructions. Furthermore, every record is assigned at least three different race and ethnicity combinations. The table shows that the race and ethnicity of: (a) *every individual from the external data is designated as (Black, non-*

*Hispanic) at least once*; (b) *at least eight different individuals are designated (AIAN, Hispanic) at least once;* (c) *and seven different individuals are designated (White, Hispanic) at least once.*

| Reconstructed Record | 1 | 2 | 3 | 4 | 5 | 6 | 7 | 8 | 9 | 10 |
|---|---|---|---|---|---|---|---|---|---|---|
| RR226 | 228 | 227 | 229 | 227 | 229 | 226 | 234 | 234 | 227 | 227 |
| RR227 | 227 | 234 | 234 | 234 | 228 | 229 | 229 | 231 | 229 | 232 |
| RR228 | 226 | 232 | 226 | 232 | 226 | 232 | 226 | 228 | 234 | 226 |
| RR229 | 234 | 233 | 232 | 229 | 230 | 227 | 227 | 227 | 231 | 228 |
| RR230 | 232 | | 233 | 226 | 231 | 228 | 235 | 232 | 228 | 230 |
| RR231 | 230 | 228 | 235 | 228 | 227 | 233 | 232 | 230 | 226 | |
| RR232 | 231 | 230 | 230 | 230 | 233 | 234 | 230 | 233 | 230 | 233 |
| RR233 | 236 | 235 | 227 | 231 | 234 | 230 | 233 | 226 | 232 | 234 |
| RR234 | 229 | 231 | 228 | 235 | 235 | 231 | 228 | | 233 | 231 |
| RR235 | 233 | 226 | 231 | 233 | | 235 | 231 | | | 229 |

Table 9. Identity from external data assigned to each reconstructed record in Block 4012 (Blank cells indicate no match was found)

| ID | 1 | 2 | 3 | 4 | 5 | 6 | 7 | 8 | 9 | 10 |
|---|---|---|---|---|---|---|---|---|---|---|
| 226 | B | W | B | B | B | W | B | B | WH | B |
| 227 | B | W | B | W | WH | A | A | A | W | W |
| 228 | W | WH | W | WH | B | B | W | B | B | A |
| 229 | W | | W | A | W | B | B | | B | W |
| 230 | WH | W | W | W | A | B | W | WH | W | B |
| 231 | W | W | W | B | B | W | W | B | A | W |
| 232 | B | B | A | B | | B | WH | B | B | B |
| 233 | W | A | B | W | W | WH | B | W | W | W |
| 234 | A | B | B | B | B | W | W | W | B | B |
| 235 | | B | WH | W | W | W | B | | | |

Table 10. Race and ethnicity assigned to individuals in Block 4012 from the 10 reconstructions (Blank cells indicate no match; W = White, not-Hispanic, WH = White, Hispanic, B = Black, A = AIAN Hispanic)

Interestingly, individuals with IDs 226, 232, and 234, who are most frequently designated as (Black, non-Hispanic), all happen to be (White, non-Hispanic). Individuals with IDs 227, 228, 230, and 233, who happen to be (Black, non-Hispanic) are *least frequently* designated as (Black, non-Hispanic). If the adversary were to rely on the reconstruction to designate race and ethnicity to the individuals in the external data (which lacks this information), the probability of a correct designation is no better than designating them randomly. If the intent of the team that designed the release of the 2010 Census tabular data was to prevent disclosure, they were extremely successful indeed!

Given that even putative identification is highly questionable, any analysis regarding confirmation of identification is entirely moot.

## Conclusions

Abowd (2021a) claims that the objective of REID analysis "a modern database reconstruction-abetted re-identification attack can reliably match a large number of 2010 census responses to the names of those respondents – a vulnerability that exposed information of at least 52 million Americans and potentially up to 179 million Americans" (p. 8). To make this claim based on a single reconstruction, *it is necessary to prove that the reconstruction was unique*. The first section of this paper definitively shows that the reconstruction is not unique.

In the absence of a unique reconstruction, the only other way for the REID team to make a claim of confirmed reidentification is to show that all (or at least most) reconstructions result in the same assignment of identity to the reconstructed records. Without this minimal level of proof, the reidentification is essentially the result of random matching. Thus far, the REID team has provided no such proof. Using *real Census data and the same analysis* as the REID team, my analysis clearly shows that *multiple reconstructions result in multiple identities being assigned to the same record in the reconstructed data*, which is all that is required to refute any claim of meaningful reidentification.

Some may consider results based on a single tract, in a single county, in a single state, or that only 10 reconstructions were performed, no better than a single reconstruction. But this would be missing the point. It is the Census Bureau which made the claim "Internal research has *conclusively proven* the fundamental vulnerabilities of the 2010 swapping methodology." (Abowd 2021c) I am simply refuting this claim by showing that, for any given track, a practically infinite number of such reconstructions exist, each reconstruction provides different results about the identity of a respondents, which casts serious doubt on this claim. It is up to the Census Bureau researchers to show either that the reconstruction was unique or, at the very least, that most reconstructions lead to the same conclusion regarding the identity of a respondent. Without such proof, claims of confirmed reidentification are highly suspect and easily refuted.

Any claims made by an adversary based on a single reconstruction can be refuted by the data administrator by issuing the following challenges:

(1) The adversary is challenged to provide the identity of the (Male, White, Hispanic, Age Group 25-29) record in Block 4012. Based on the first reconstruction, the adversary identifies this record as belonging to the individual with ID 230 in the external data. Using the same data, the data administrator counters by showing that, based on the remaining nine reconstructions, that this record could also belong to individuals with IDs (228, 235, 227, 233, 232, 230, 226) or not identified at all (reconstruction 10).

(2) The adversary is challenged to identify the (Black, non-Hispanic) individuals in Block 4012. Based on the first reconstruction, the adversary identifies individuals with ID (226, 227, 232) in the external data as being (Black, non-Hispanic) since an age match was not found for one reconstructed record. Using the same data, the data administrator counters with the remaining nine reconstructions to show that any of the 10 individuals could be identified as (Black, non-Hispanic).

The interesting fact is that, *to prevent any disclosure, the administrator never actually confirms or denies the adversary's claim*. The administrator simply shows that there exist reconstructions that present alternative solutions that refute the adversary's claim. Faced with these facts, the adversary has no recourse but to acknowledge that the claims of reidentification cannot be substantiated. As the data administrator, it would be the Census Bureau's *duty* to challenge the claims made by the REID team, which they have not. I am doing so on behalf of the public.

**Acknowledgement**: I would like to thank Margo Anderson, Jane Bambauer, and Connie Citro for their helpful comments and suggestions, and Carolyn Jensen, Business Communication Center, Price College of Business, for her editorial assistance.

## References

Abowd, J.M. (2018). *Staring Down the Database Reconstruction Theorem*, Joint Statistical Meeting, Vancouver B.C., Canada.

Abowd, J. (2021a). *Declaration of John Abowd, State of Alabama v. United States Department of Commerce*. Case No. 3:21-CV-211-RAH-ECM-KCN. http://vhdshf2oms2wcnsvk7sdv3so.blob.core.windows.net/thearp-media/documents/Declaration_of_John_M._Abowd.pdf (downloaded January 14, 2022).

Abowd, J. (2021b). *Supplemental Declaration of John M. Abowd, State of Alabama v. United States Department of Commerce*. Case No. 3:21-CV-211-RAH-ECM-KCN. https://www.brennancenter.org/sites/default/files/2021-06/M.D.%20Ala.%2021-cv-00211%20dckt%2000000116_001%20filed%202021-04-26%20Abowd%20declaration.pdf (downloaded January 14, 2022).

Abowd, J. (2021c) *Second Declaration of John M. Abowd, Fair Lines America Foundation Inc. V. United States Depart of Commerce and United States Bureau of the Census.* Case No. Civ. A. No. 1:21-cv-01361 (ABJ) https://www2.census.gov/about/policies/foia/records/disclosure-avoidance/abowd-fair-lines-v-commerce-second-declaration.pdf (downloaded March 15, 2022).

Rastogi, S. and O'Hara, A. (2012) *2010 Census Match Study Report*, 2010 CENSUS PLANNING MEMO NO. 247, United States Census Bureau, Washington DC. https://www.census.gov/library/publications/2012/dec/2010_cpex_247.html (downloaded November 27, 2021)

Ruggles, S. and Van Riper, D. (2021) The Role of Chance in the Census Bureau Database Reconstruction Experiment, *Population Research and Policy Review* (published online). https://link.springer.com/article/10.1007%2Fs11113-021-09674-3 (downloaded November 27, 2021).